\documentclass[aps,prx,twocolumn,superscriptaddress,letterpaper]{revtex4-1}

\usepackage{graphicx}
\usepackage{bm}
\usepackage{amssymb}
\usepackage{color}
\usepackage{amsmath}
\usepackage{ulem}
\usepackage{siunitx}

\begin{document}

\title{Vortex states in an acoustic Weyl crystal with a topological lattice defect}


\author{Qiang Wang}
\affiliation{Division of Physics and Applied Physics, School of Physical and Mathematical Sciences, Nanyang Technological University,
Singapore 637371, Singapore}

\author{Yong Ge}
\affiliation{Research Center of Fluid Machinery Engineering and Technology, School of Physics and Electronic Engineering, Jiangsu University, Zhenjiang 212013, China.}

\author{Hong-xiang Sun}
\affiliation{Research Center of Fluid Machinery Engineering and Technology, School of Physics and Electronic Engineering, Jiangsu University, Zhenjiang 212013, China.}

\author{Haoran Xue}
\affiliation{Division of Physics and Applied Physics, School of Physical and Mathematical Sciences, Nanyang Technological University,
Singapore 637371, Singapore}

\author{Ding Jia}
\affiliation{Research Center of Fluid Machinery Engineering and Technology, School of Physics and Electronic Engineering, Jiangsu University, Zhenjiang 212013, China.}

\author{Yi-jun Guan}
\affiliation{Research Center of Fluid Machinery Engineering and Technology, School of Physics and Electronic Engineering, Jiangsu University, Zhenjiang 212013, China.}

\author{Shou-qi Yuan}
\email{shouqiy@ujs.edu.cn}
\affiliation{Research Center of Fluid Machinery Engineering and Technology, School of Physics and Electronic Engineering, Jiangsu University, Zhenjiang 212013, China.}

\author{Baile Zhang}
\email{blzhang@ntu.edu.sg}
\affiliation{Division of Physics and Applied Physics, School of Physical and Mathematical Sciences, Nanyang Technological University,
Singapore 637371, Singapore}
\affiliation{Centre for Disruptive Photonic Technologies, Nanyang Technological University, Singapore, 637371, Singapore}

\author{Y. D. Chong}
\email{yidong@ntu.edu.sg}
\affiliation{Division of Physics and Applied Physics, School of Physical and Mathematical Sciences, Nanyang Technological University,
Singapore 637371, Singapore}

\affiliation{Centre for Disruptive Photonic Technologies, Nanyang Technological University, Singapore, 637371, Singapore}

\begin{abstract}
We design and implement a three dimensional acoustic Weyl metamaterial hosting robust modes bound to a one-dimensional topological lattice defect.  The modes are related to topological features of the bulk bands, and carry nonzero orbital angular momentum locked to the direction of propagation.  They span a range of axial wavenumbers defined by the projections of two bulk Weyl points to a one-dimensional subspace, in a manner analogous to the formation of Fermi arc surface states.  We use acoustic experiments to probe their dispersion relation, orbital angular momentum locked waveguiding, and ability to emit acoustic vortices into free space.  These results point to new possibilities for creating and exploiting topological modes in three-dimensional structures through the interplay between band topology in momentum space and topological lattice defects in real space.
\end{abstract}

\maketitle

\section{Introduction}

Topological lattice defects (TLDs) are crystallinity-breaking defects in lattices that cannot be eliminated by local changes to the lattice morphology, due to their nontrivial real-space topology \cite{Mermin1979}.  Although they give rise to numerous important physical effects in their own right \cite{Kosterlitz2017}, TLDs can have especially interesting consequences in materials with topologically nontrivial bandstructures \cite{jackiw1981, Lammert2000, Ran2009, Teo2010, Juri2012, Ruegg2013}.  For instance, Ran \textit{et al.}~have shown theoretically that introducing a screw dislocation into a three dimensional (3D) topological band insulator induces the formation of one dimensional (1D) helical defect modes, which are protected by the interplay between the Burgers vector of the defect and the topology of the bulk bandstructure \cite{Ran2009}.  Aside from topological band insulators \cite{Ran2009, Slager2014, Slager2015, Slager2019}, other topological phases are predicted to have their own unique interactions with TLDs, including Weyl semimetals, topological crystalline insulators, and higher-order topological insulators \cite{sumiyoshi2016, liu2017, soto2020, vanMiert2018, LiHughes2020, queiroz2019}.  TLD-induced modes provide a way to probe bandstructure topology independent of standard bulk-boundary correspondences \cite{Teo2010, Juri2012, Ruegg2013, Slager2015, soto2020, LiHughes2020}, and may give rise to exotic material properties such as anomalous torsional effects \cite{sumiyoshi2016}. Experimental confirmations have, however, been hampered by the difficulty of accessing TLDs in real topological materials \cite{Yazyev2010, Huang2011, Hamasaki2020}.  Recently, various groups have turned to classical wave metamaterials \cite{lin2018, li2018, wang2020, liu2021, peterson2021} to perform the experimental studies of the interplay between TLDs and topological bandstructures, including the demonstration of topologically-aided trapping of light on a dislocation \cite{li2018}, robust valley Hall-like waveguiding along disclination lines \cite{wang2020}, and defect-induced fractional modes \cite{liu2021, peterson2021}.  The preceding studies have all been based on two dimensional (2D) lattices; 3D lattices with TLD-induced topological modes have thus far only been investigated theoretically.

Here, we design and experimentally demonstrate a 3D acoustic metamaterial that hosts topological modes induced by the presence of a TLD.  Without the TLD, the bulk metamaterial forms a Weyl crystal, whose 3D bandstructure contains topologically nontrivial degeneracies called Weyl points \cite{wan2011, yang2014, liu2014, Lu2015, fang2015, Xu2015, lv2015, bian2016, noh2017, wang2017, wu2018, armitage2018, li2018weyl, yan2018, xiao2020}.  Weyl crystals are known to exhibit, along their 2D external surfaces, Fermi arc states that are protected by the topology of the Weyl points \cite{lv2015, Xu2015}.  The introduction of the defect generates a family of modes localised to the line of the TLD (in real space).  Moreover, in a manner analogous to the formation of regular Fermi arcs, the modes span the projections of two Weyl points of opposite topological charge in the axial momentum space $k_z$.  Hence, we refer to them as TLD-induced Fermi arc (TIFA) modes.  The TIFA mode for each $k_z$ can be interpreted as a 2D bound state generated by a strongly localised pseudo-magnetic flux associated with the TLD, in accordance with earlier theoretical predictions about disclinations in 2D topological materials \cite{Ruegg2013}.  Hence, the TIFA modes arise from the interplay between the TLD and the 3D Weyl bandstructure.

Recently, the discovery of higher-order topological materials \cite{benalcazar2017} has led to the idea of higher-order Weyl and Weyl-like phases \cite{Lin2018HO, Bitan2019, ifmmode2019, Wang2020HO, ghorashi2020, luo2020, wieder2020, Wu2020, wei2021}, which can host ``higher-order Fermi arcs'' \cite{Wang2020HO, ghorashi2020,luo2020, wei2021}.  Like the TIFA modes discussed in this paper, higher-order Fermi arc modes are one-dimensional, but they arise from a completely different mechanism involving higher-order topological indices \cite{Wang2020HO, ghorashi2020, wei2021}.  Moreover, they lie along external hinges, whereas the TIFA modes are localised to the line of the TLD, embedded inside a 3D bulk.

\begin{figure*}
\centering
\includegraphics[width=0.95\textwidth]{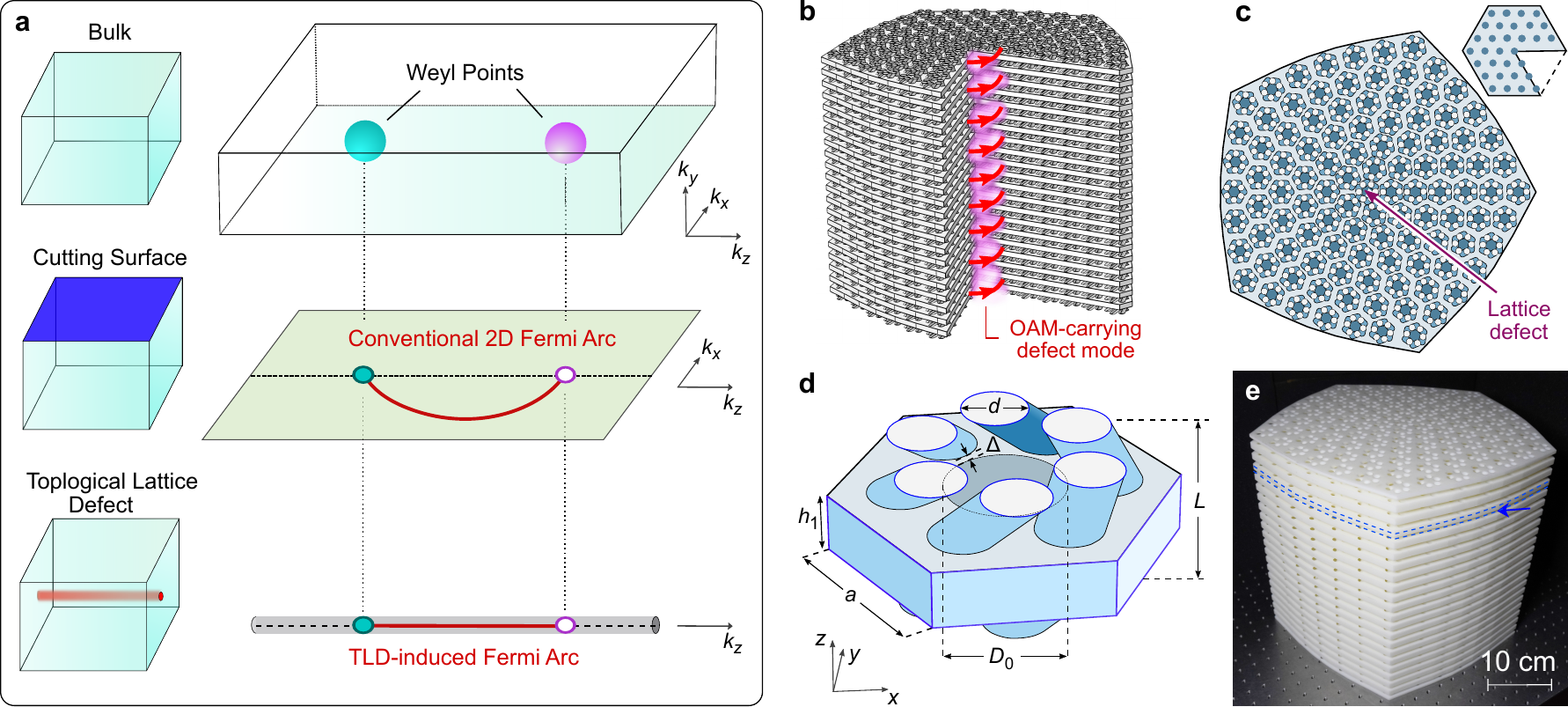}
\caption{\textbf{Weyl acoustic structure with a topological lattice defect (TLD)}.  \textbf{a}, Conceptual illustration of TLD-induced Fermi arc (TIFA) mode formation.  Top row: a bulk crystal (left) hosts two Weyl points with opposite topological charges in 3D momentum space (right).  Middle row: truncating the crystal (left) creates a Fermi arc extending between the projections of the Weyl points in the 2D surface momentum space (right). Bottom row: adding a TLD (left) results in TIFA modes that extend between the projections of the Weyl points into the 1D momentum space (right).  \textbf{b},~Schematic of an acoustic lattice made of chiral layers with a central TLD, stacked along $z$.  A section is omitted to show the internal structure.  A TIFA mode carrying nonzero orbital angular momentum (OAM) propagates along the TLD (pink region). \textbf{c}, Top-down schematic of one layer.  The TLD is generated by deleting a $\pi/3$ wedge from a triangular lattice (inset).  \textbf{d},~Close-up view of one unit cell.  The periodicity in $z$ is $L = 1.8\,\textrm{cm}$.  Each unit cell consists of an air-filled sheet of thickness $h_1=0.8\,\textrm{cm}$, with a central solid rod of diameter $D_0=1.6\,\textrm{cm}$ ringed by six skewed air-filled tubes of diameter $d=0.9\,\textrm{cm}$ in a hexagonal arrangement.  The tubes advance by $\pi/6$ around the hexagon between layers.  The spacing between the central rod and the tubes is $\Delta=0.1\,\textrm{cm}$, and the side length of the hexagonal cell is $a=4/\sqrt{3}\,\textrm{cm}$.  All other regions are solid resin.  \textbf{e},~Photograph of the experimental sample with 21 layers.  Blue dashes indicate one of the layers, and the arrow indicates one of the gaps for inserting acoustic probes. }
\label{fig:sketch}
\end{figure*}

The TIFA modes carry nonzero orbital angular momentum (OAM), locked to their propagation direction. For each $k_z$, the sign of the OAM depends on the Chern number of the 2D projected band structure, and matches the chirality of the robust localised state predicted to occur in a Chern insulator on a 2D surface with singular curvature \cite{WenXG1992,Parrikar2014, Andreas2013, schine2016} (a prediction that has never previously been verified in an experiment \cite{Ruegg2013, Can2016, biswas2016}).  To our knowledge, this is also the first demonstration of a 3D topology-induced mode carrying nonzero OAM.  Classical waves with nonzero OAM have a variety of emerging applications including vortex traps and rotors \cite{skeldon2008, baresch2016} and OAM-encoded communications \cite{shi2017}.  Although chiral structures have previously been studied for the purposes of OAM waveguiding, those waveguides support multiple OAM modes with different propagation constants \cite{wong2012}; by contrast, the present topological waveguide supports, for each $k_z$, a single robust TIFA mode with nonzero OAM.

\section{Lattice Design}

The emergence of a TLD-induced Fermi arc (TIFA) mode is conceptually illustrated in  Fig.~\ref{fig:sketch}\textbf{a}.  In a Weyl semimetal, topologically-charged Weyl points in the 3D bulk imply the existence of Fermi arc modes on 2D external surfaces of the crystal. In the 2D surface momentum space, each Fermi arc extends between the projections of two oppositely-charged Weyl points. The introduction of a TLD into the Weyl crystal breaks translational symmetry in the $x$-$y$ plane while maintaining it along $z$, and generates TIFA modes that are spatially localised to the 1D string formed by the TLD.  Viewed from momentum space, the TIFA modes extend between the projections of the two Weyl points into the residual 1D momentum space $k_z$.

To realise this phenomenon, we designed and fabricated a 3D acoustic crystal formed by chirally structured layers stacked along $z$, as shown in Fig.~\ref{fig:sketch}\textbf{b}--\textbf{e}.  Without the TLD, an $x$-$y$ cross section of the structure would form a triangular lattice.  The TLD is introduced by a ``cut-and-glue'' procedure in which a $\pi/3$ wedge is deleted (Fig.~\ref{fig:sketch}\textbf{c} inset) and the edges are reattached by deforming the rest of the lattice (see Methods).  The experimental sample is formed by stacking 3D-printed structures, with a total of 21 layers (see Methods); a photograph is shown in Fig.~\ref{fig:sketch}\textbf{e}.


\begin{figure} 
\centering
\includegraphics[width=0.49\textwidth]{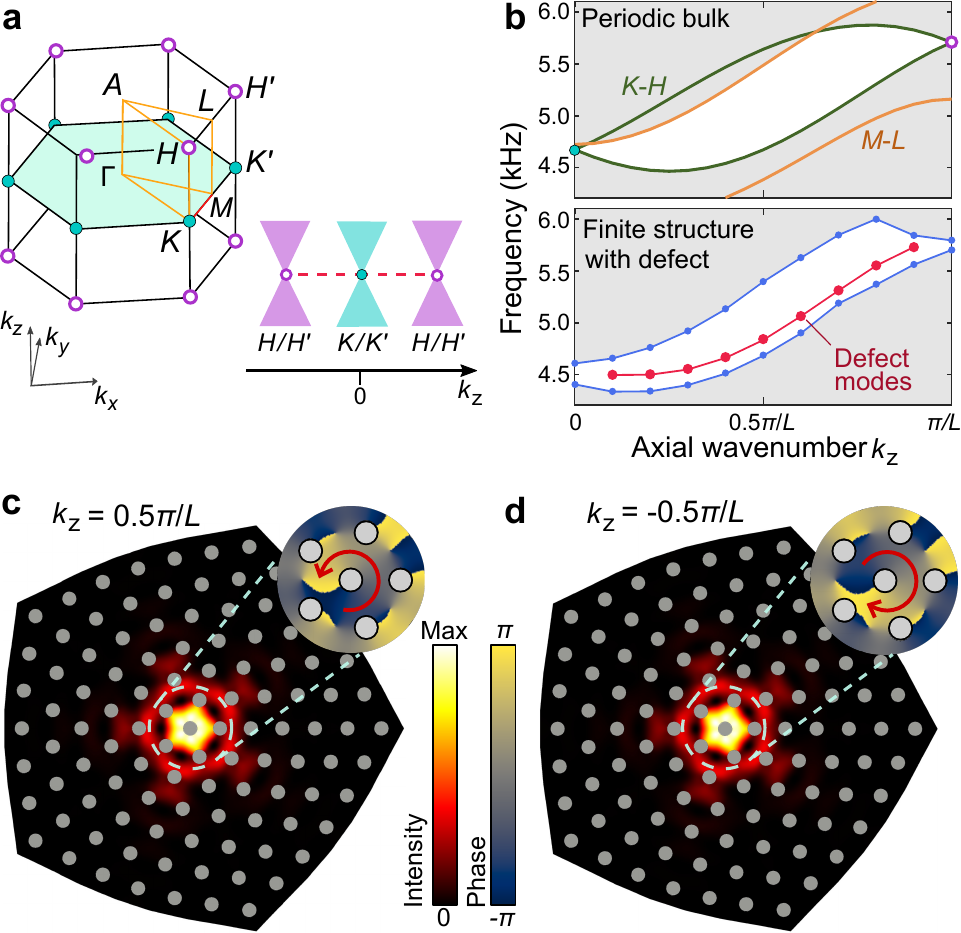}
\caption{\textbf{Numerical characterisation of the TIFA modes}. \textbf{a}, Left panel: 3D Brillouin zone of the acoustic crystal without a TLD.  Weyl points occur at $K$ and $K'$ with topological charge +1 (cyan dots), and at $H$ and $H'$ with charge -1 (magenta dots).  Right panel: projection of the Weyl points onto $k_z$, with red dashes indicating the TIFA modes. \textbf{b}, Numerical bandstructures.  Upper plot: bands along $K$-$H$ and $M$-$L$ for the periodic bulk.  Lower plot: bands for a structure with a TLD (with the same cross sectional profile as in Fig.~\ref{fig:sketch}\textbf{b}--\textbf{e}) and periodicity $L$ along $z$; band edge (in-gap) modes are plotted in blue (red).  In-gap regions are shown in white.  All bands, including the TIFA modes, are symmetric around $k_z = 0$; only the $k_z > 0$ range is plotted here.   \textbf{c},\,\textbf{d},~Calculated acoustic pressure intensity distributions in the $x$-$y$ plane, with $z$ at the midpoint of the structure's central air sheet, for the TIFA modes at $k_z=\pi/2L$ (\textbf{c}) and $k_z=-\pi/2L$ (\textbf{d}).  Both modes have frequency $4.844\,\textrm{kHz}$.  The gray circles are the solid rods passing through the air sheet.  Insets: phase distribution of acoustic pressure near the TLD, showing that the two modes have opposite OAM.}
\label{fig:sim}
\end{figure}

\begin{figure*}
\centering
\includegraphics[width=\textwidth]{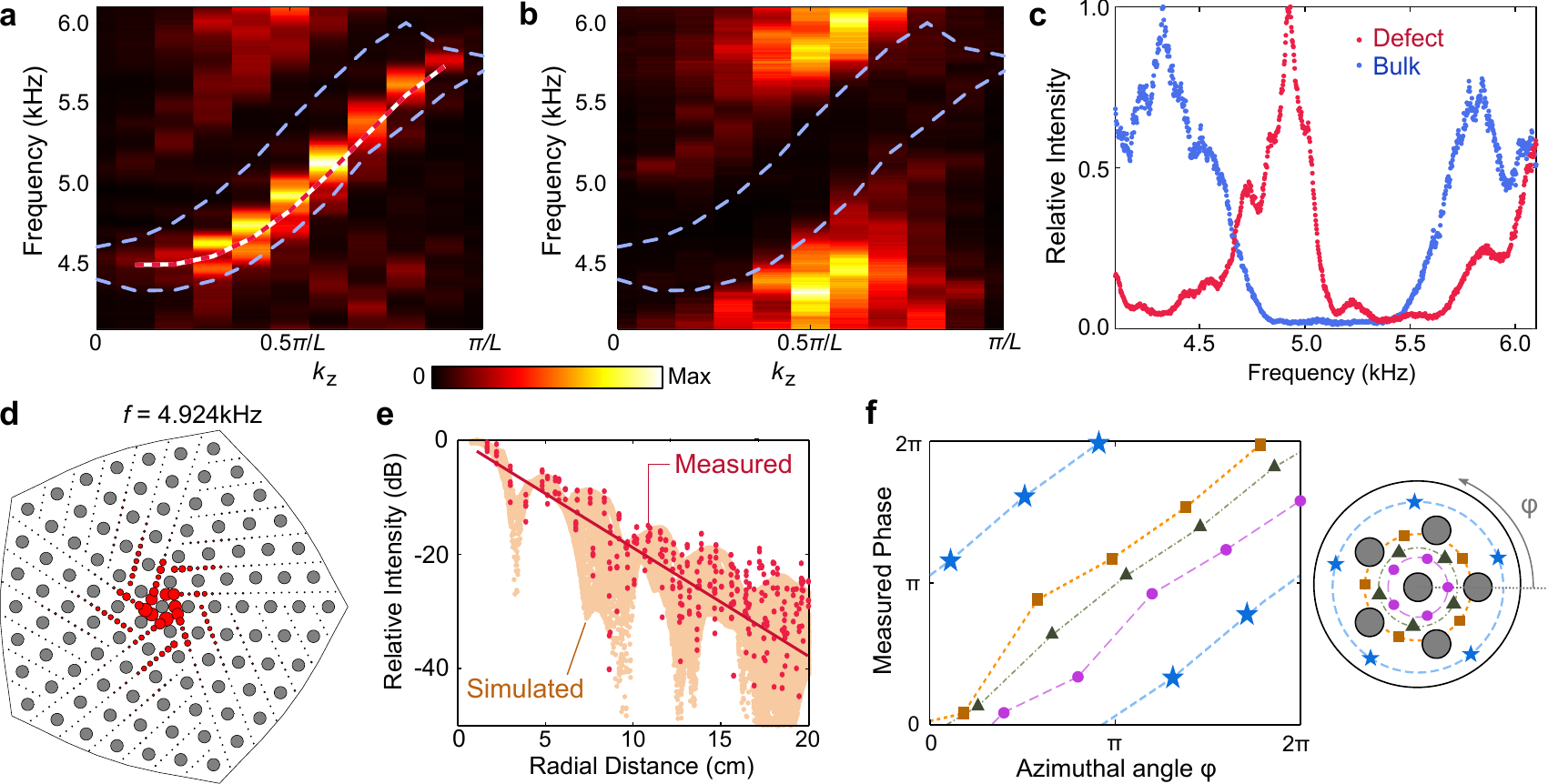}
\caption{\textbf{Experimental observation of TIFA modes}.  \textbf{a}, Measured spectral intensity of the TIFA modes.  With an acoustic source in the bottom layer near the TLD, the acoustic pressure readings are Fourier transformed in $z$, and the mean in-plane intensity within  $\textrm{5 cm}$ of the TLD is plotted.  Red-and-white dashes show the numerically obtained TIFA mode dispersion curve, and blue dashes show the numerically obtained band edges. \textbf{b},~Spectral intensity of bulk modes, obtained by placing the source and probe $14\,\textrm{cm}$ and $10\,\textrm{cm}$ away from the TLD, respectively.  \textbf{c},~Defect (red) and bulk (blue) spectral intensities at $k_z=\pi/2L$, with each curve normalised to its maximum value.  \textbf{d}, Measured in-plane intensity distribution at $k_z=\pi/2L$ and frequency $4.924\,\textrm{kHz}$.  Each red circle is centered at a measurement point, and its area is proportional to the squared magnitude of the acoustic pressure. \textbf{e}, Semi-logarithmic plot of intensity versus radial distance from the TLD for the mode in \textbf{d}.  Red dots show the experimental results, taken at different azimuthal angles, and orange dots show the profile of the numerically obtained TIFA eigenmode.  \textbf{f}, Left panel: measured phase signal versus azimuthal angle $\varphi$ at different radial distances near the TLD.  Right panel: schematic of the measurement positions.  In \textbf{d} and \textbf{f}, the gray circles indicate the solid rods. }
\label{fig:exp-dis}
\end{figure*}

The 3D Brillouin zone of the acoustic crystal, in the absence of the TLD, is depicted in the left panel of Fig.~\ref{fig:sim}\textbf{a}.  Weyl points exist at $K$ and $K'$ ($H$ and $H'$), with topological charge +1 (-1) \cite{xiao2015, peri2019, li2018weyl}; for details, refer to Supplementary Section I.  Consider the Weyl point at $K$ or $K'$ (the analysis for $H$ and $H'$ is similar).  In its vicinity, the wavefunctions are governed by the effective Hamiltonian
\begin{equation}
  \mathcal{H} = -i  \big( \tau_z \sigma_x \partial_x + \sigma_y \partial_y \big)
  + k_z \tau_z\sigma_z,
  \label{H0}
\end{equation}
where $\tau_i$ ($\sigma_i$) denotes valley (sublattice) Pauli matrices, we have rescaled each spatial coordinate so that the group velocity is unity, and $k_z$ is the wavenumber in the $z$ direction.  With the introduction of the TLD, $k_z$ remains a good quantum number; in the $x$-$y$ plane, the distortion introduced by the TLD can be modelled as a matrix-valued gauge field \cite{Ruegg2013} that mixes the valleys (i.e., $K$ with $K'$ and $H$ with $H'$). The effective Hamiltonian can be brought back into block-diagonal form by a unitary transformation \cite{Ruegg2013}, whereby the Hamiltonian for each block has the form of Eq.~\eqref{H0} but modified by
\begin{equation}
  \tau_z \rightarrow \tau', \;\; \nabla \rightarrow \nabla + i \tau'\mathbf{A}, \;\; \mathbf{A} = (4\Omega r)^{-1} \mathbf{e}_\theta,
\label{blocks}
\end{equation}
where $\tau' = \pm 1$ is the block index, $r$ is the radial coordinate and $\mathbf{e}_\theta$ is the azimuthal unit vector in the unfolded space, and the factor $\Omega = 5/6$ is the number of undeleted wedges.  Unlike previously studied strain-induced pseudo-magnetic fluxes in Weyl semimetals \cite{jia2019, peri2019, ilan2019}, the pseudo-magnetic flux here is strongly localised \cite{Ruegg2013, Gonzalez1993}.  Moreover, unlike previous studies of pseudo-magnetic fluxes generated by screw dislocations, the pseudo-magnetic flux is $k_z$-independent \cite{Ran2009, sumiyoshi2016}.

Hence, viewed from 2D, the pseudo-magnetic flux induces topologically protected chiral defect states.  For each $k_z > 0$, one can show \cite{Ran2009, Ruegg2013} that there is a single bound solution (among the two Weyl Hamiltonians) localised at $r = 0$.  This remains true even when $k_z$ is non-perturbative.  For fixed $k_z$, the lattice in the absence of the TLD maps to a 2D Chern insulator whose Chern numbers switch sign with $k_z$ (the gap closes at $0$ and $\pm \pi/L$); upon introducing the TLD via the cut-and-glue construction, one of the two sub-blocks in the effective Hamiltonian ($\tau' = 1$ for $0 < k_z < \pi/L$, and $\tau' = -1$ for $-\pi/L < k_z < 0$) exhibits a solution that is localised to the TLD\cite{Ruegg2013}.  As we vary $k_z$, this family of solutions spans the projections of the Weyl points at $K$($K'$) and $H$($H'$). Note that the overall acoustic structure preserves time-reversal symmetry ($T$), but the individual Hamiltonian sub-blocks effectively break $T$; the defect mode at $-k_z$ thus serves as the time-reversed counterpart of the defect mode at $k_z$, with opposite chirality.  For further details, refer to Supplementary Section II.

The upper panel of Fig.~\ref{fig:sim}\textbf{b} shows the numerically computed acoustic band diagram for the TLD-free bulk structure, projected onto $k_z$.  The relevant bands along $K$-$H$ ($M$-$L$) are plotted in green (orange), and the gap region is shown in white.  The lower panel of Fig.~\ref{fig:sim}\textbf{b} shows the corresponding band diagram for a structure with a TLD, which is periodic along $z$ and has the same $x$-$y$ profile as the experimental sample (Fig.~\ref{fig:sketch}\textbf{b}--\textbf{e}).  These numerical results reveal the existence of TIFA modes, plotted in red, which occupy the gap and span almost the entire $k_z$ range.  (Near $k_z = 0$ and $k_z = \pi/L$, they are difficult to distinguish from bulk modes due to finite-size effects.)

In Fig.~\ref{fig:sim}\textbf{c},~\textbf{d}, we show the mode distributions for the TIFA modes at $k_z = \pm 0.5 \pi/L$.  The modes are strongly localised to the center of the TLD; their intensity profiles are identical since the two modes map to each other under time reversal.  The phase distributions (inset) reveal that the $k_z > 0$ ($k_z < 0$) TIFA mode has winding number +1 (-1).  This winding number is tied to the Chern number of the 2D projected band structure for fixed $k_z$.  The fact that the TIFA modes carry nonzero OAM, locked to the propagation direction, distinguishes them from previously studied topological defect modes \cite{Gao2019sonic, Menssen2020, gao2020, noh2020} and hinge modes \cite{Wang2020HO, ghorashi2020, wei2021} that have zero OAM.  Moreover, we have verified numerically that the TIFA modes' localisation and OAM are robust to in-plane disorder, consistent with their topological origin (see Supplementary Section III).

\section{Experimental Results}

We performed a variety of experiments to characterise the TIFA modes in the fabricated structure.  First, we investigated their dispersion curve by threading an acoustic source into the bottom layer of the sample, near the center of the TLD.  A probe is inserted into the other 20 layers in turn, via the central air sheet in each layer, as indicated by the blue arrow in Fig.~\ref{fig:sketch}\textbf{d}.  The acoustic pressure, measured close to the center of the TLD, is Fourier transformed to obtain the spectral plot shown in Fig.~\ref{fig:exp-dis}\textbf{a}.  The overlaid red dashes are the numerically obtained TIFA mode dispersion curve (Fig.~\ref{fig:sim}\textbf{b}), which closely matches the intensity peaks in the experimental results.  We then repositioned the source and probe away from the TLD, obtaining in the spectrum shown in Fig.~\ref{fig:exp-dis}\textbf{b}; this matches the bulk spectrum obtained numerically, with the spectral intensities peaking in the bulk bands.  For details about the source and probe positions, see Supplementary Section IV.

The acoustic pressure intensity at $k_z=\pi/2L$ is plotted versus frequency in Fig.~\ref{fig:exp-dis}\textbf{c}.  A narrow peak corresponding to the TIFA modes is clearly observable within the bulk gap, with only a small frequency shift of $80\,\textrm{Hz}$ relative to the numerically predicted eigenfrequency.  For excitation near the TLD, the measured intensity distribution at frequency $f = 4.924\,\textrm{kHz}$ is plotted in Fig.~\ref{fig:exp-dis}\textbf{d}, showing strong localisation around the TLD.  The radial dependence of the intensity distribution is plotted in Fig.~\ref{fig:exp-dis}\textbf{e} (note that the apparent irregularity arises from the fact that the measurement points lie at different azimuthal angles).  The measurement data is in good agreement with the numerically obtained TIFA mode profile.  From a linear least squares fit of the semi-logarithmic plot, using measurement data up to a radial distance of $12\,\textrm{cm}$, we find a localisation length of $2.38\,\textrm{cm}$, which is on the order of the mean distance between unit cells (i.e., the approximate lattice constant).

Fig.~\ref{fig:exp-dis}\textbf{f} plots the phase of the measured acoustic pressure signal versus azimuthal angle for $k_z=\pi/2L$ and $f = 4.924\,\textrm{kHz}$.  The different data series in this plot correspond to measurement points at different radial distances.  The phase is observed to wind by $+2\pi$ during a counterclockwise loop encircling the TLD, consistent with the numerically obtained eigenmode (Fig.~\ref{fig:sim}\textbf{c}), which implies that the TIFA mode has OAM of +1.

\begin{figure}
\centering
\includegraphics[width=0.9\columnwidth]{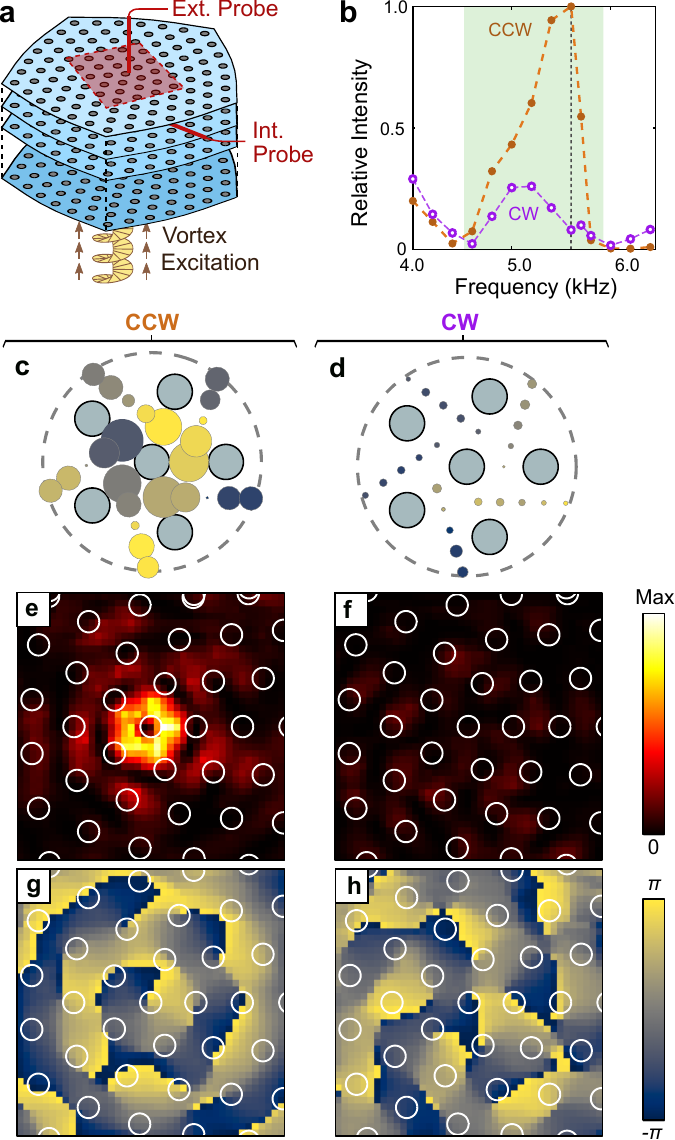}
\caption{\textbf{Selective excitation by a vortex source}.  \textbf{a}, Schematic of the experiment.  A counterclockwise (CCW) or clockwise (CW) acoustic vortex is incident on the bottom of the sample, centered on the TLD.  An external probe, $2\,\textrm{mm}$ above the top surface, sweeps over a $20\,\textrm{cm}\times 20\,\textrm{cm}$ area ($0.5\,\textrm{cm}$ step size).  An internal probe is inserted into the top layer. \textbf{b}, Normalised acoustic pressure intensity versus frequency measured in the top layer by the internal probe for a CCW (purple dots) and CW (orange dots) vortex source.  The green region indicates the frequency range hosting TIFA modes.  \textbf{c}, \textbf{d}, Acoustic pressure distribution in the top layer for a CCW (\textbf{g}) and CW (\textbf{h}) vortex source at $5.6\,\textrm{kHz}$ (vertical dotted line in \textbf{b}).  The area and color of each circle correspond to intensity and phase respectively.   \textbf{e},~\textbf{h}, Intensity distributions (\textbf{e}, \textbf{f}) and phase distributions (\textbf{g}, \textbf{h}) measured by the external probe for a CCW (\textbf{c}) and CW (\textbf{d}) vortex source.  The gray circles in \textbf{c}--\textbf{d} and white circles in \textbf{e}--\textbf{h} indicate the structural rods. }
\label{fig:outsidesweep}
\end{figure}

To demonstrate the physical significance of the OAM carried by the TIFA modes, we studied their coupling to external acoustic vortices.  The experimental setup is shown in Fig.~\ref{fig:outsidesweep}\textbf{a}.  The vortex wave is generated in a cylindrical waveguide of radius $1.7\,\textrm{cm}$, attached to the bottom layer of the sample at the center of the TLD.  Fig.~\ref{fig:outsidesweep}\textbf{b} shows the acoustic pressure intensity measured in the top layer, on the opposite side of the sample from the source.  This intensity is obtained by averaging over points closest to the TLD, and dividing by the averaged intensity in the bottom layer to normalise away the frequency dependence of the source.  For a counterclockwise (CCW) vortex source, a strong peak is observed within the range of frequencies where TIFA modes are predicted to exist.  For a clockwise (CW) vortex source, the intensity is low (the non-vanishing intensity is likely due to finite-size effects).  Fig.~\ref{fig:outsidesweep}\textbf{c}--\textbf{d} shows the intensity and phase distributions measured in the top layer at $5.6\,\textrm{kHz}$, which confirm that the TIFA modes are preferentially excited by the CCW vortex.

After the TIFA modes have passed through the structure, they emit an acoustic vortex into free space at the far surface.  In Fig.~\ref{fig:outsidesweep}\textbf{e}--\textbf{h}, we show the intensity and phase distributions measured by an external acoustic probe positioned $2\,\textrm{mm}$ above the top surface of the sample.  For a CCW vortex source in the bottom layer, a CCW vortex is emitted from the top layer, at the position of the TLD; for a CW vortex source, the emission is negligible due to the TIFA modes not being excited.  For frequencies outside the range of the TIFA modes, the CW and CCW vortices both produce negligible emission from the top layer (see Supplementary Section IV).

\section{Discussion}
We have experimentally realised a 3D acoustic structure hosting localised topological modes induced by a topological lattice defect, which we have called TIFA modes.  In real space, the modes lie along a one dimensional line formed by the defect, embedded within the bulk; in momentum space, they connect the projections of the Weyl points in the defect-free crystal, and hence span the one dimensional Brillouin zone.  This is, to our knowledge, the first experimental demonstration of a defect-induced topological mode in any 3D system.  For each momentum space slice ($k_z$), the system maps onto a 2D Chern insulator trapped on a surface with singular curvature.  Theoretical studies had previously shown that such a system hosts a robust localised defect mode tied to the Chern number of the 2D bulk bandstructure \cite{Ruegg2013, Can2016, biswas2016}.

The TIFA modes carry nonzero OAM, locked to their propagation direction.  This is a striking feature not possessed by topological defect modes based on other similar schemes; for example, the localised topological modes of 2D Kekul\'e lattices carry zero winding number \cite{Gao2019sonic, Menssen2020, noh2020, gao2020}.  Our sample therefore serves as an OAM-locked acoustic waveguide, one whose operating principles are very different from the chiral acoustic emitters \cite{ealo2011, jiang2016b} and metasurfaces \cite{jiang2016, Fu2020} studied in previous works.  This design may be useful for applications of acoustic vortices, such as acoustic traps and rotors\cite{skeldon2008, baresch2016} and OAM-encoded communications \cite{shi2017}.  Similar designs could be used to realise TIFA modes in photonics, based on three dimensional photonic crystals \cite{Lu2015} or laser-written waveguide arrays \cite{noh2017}.

Finally, our work opens the door for further investigations into the numerous other effects of lattice defects in topological materials.  Many interesting phenomena in this area have been proposed theoretically but have not thus far been observed, including torsional chiral magnetic effects in Weyl semimetals and 1D helical defect modes in 3D weak topological insulators\cite{Ran2009, sumiyoshi2016, soto2020}.

\vskip 0.1in
\noindent \textbf{Acknowledgements}\;
Q.~W.~,  H.~X.~, B.~Z.~and Y.~C.~ acknowledge support from Singapore MOE Academic Research Fund Tier 3 Grant MOE2016-T3-1-006, Tier 1 Grants RG187/18, and Tier 2 Grant MOE2019-T2-2-085. Y.~G., H.-X.~S., D.~J., Y.-J.~G. and Y.-Q.~S. acknowledge  support from  National Natural Science Foundation of China under Grants No. 11774137 and 51779107,  National Key R\&D Program Project (No.~2020YFC1512403 and 2020YFC1512400) and the State Key Laboratory of Acoustics, Chinese Academy of Science under Grant No. SKLA202016.

\vskip 0.1in
\noindent \textbf{Author contributions}\;
Q.~W.~and Y.~G.~contributed equally to this work.  Q.~W., B.~Z.~and Y.~C.~conceived the idea. Q.~W.~designed the acoustic structures and performed the numerical simulations.  Q.~W.~, H.~X.~, and H.-X.~S.~ designed the experiments and fabricated the sample. Y.~G., Y.-J.~G., D.~J., and H.-X.~S.~conducted the measurements.  Y.-Q.~S., B.~Z.~and Y.~C.~supervised the project.  All authors contributed extensively to the interpretation of the results and the writing of the paper.

\bibliography{citepaper.bib}

\begin{widetext}
\newpage

\makeatletter 
\renewcommand{\theequation}{S\arabic{equation}}
\makeatother
\setcounter{equation}{0}

\makeatletter 
\renewcommand{\thefigure}{S\@arabic\c@figure}
\makeatother
\setcounter{figure}{0}

\makeatletter 
\renewcommand{\thesection}{Section \arabic{section}}
\makeatother
\setcounter{section}{0}

\begin{center}
  {\small \textbf{Supplemental Materials for}}\\
  {\large ``Vortex states in an acoustic Weyl crystal with a topological lattice defect''}\\
  {\small Q.~Wang, Y.~Ge, H.-X. Sun, H.~Xue, D.~Jia, Y.-J.~Guan, S.-Q~Yuan, B.~Zhang, and Y.~D.~Chong}
\end{center}
 
\section{Numerical simulation for the Charge of Weyl points}

In this section, we describe the numerical determination of the charges of the Weyl points for the 3D acoustic structure in the absence of a topological lattice defect (TLD).  We adopt the method described by Soluyanov \textit{et al.}~\cite{soluyanov2015}.  A spherical surface is chosen to enclose one Weyl point, and we calculate the Berry phase around loops of constant latitudinal angle $\theta$, as shown in Fig.~\ref{fig:charge}\textbf{a}.  The evolution of the Berry phase as a function of $\theta$ is plotted in Fig.~\ref{fig:charge}\textbf{b},\textbf{c} for the Weyl points at $K$ and $H$ respectively. The Berry phase for the lower bands (red circles) exhibits a $2\pi$ ($-2\pi$) change, implying that the charge for the two Weyl points are +1 and -1. The eigenfunctions used to calculate the Berry phase are obtained from \textit{ab initio} acoustic simulations performed with Comsol Multiphysics.

\begin{figure}[b]
\centering
\includegraphics[width=0.9\columnwidth]{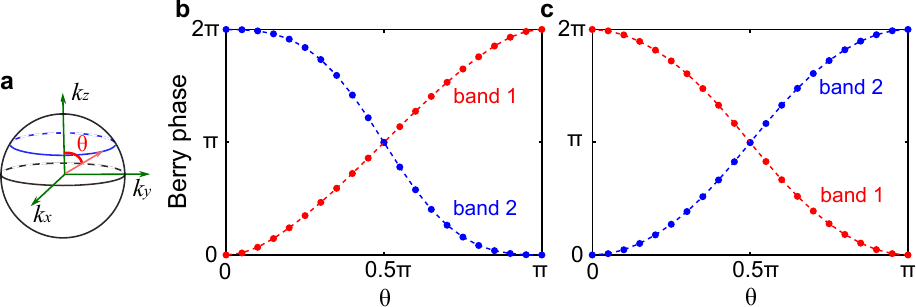}
\caption{Numerical calculation of Weyl point charges.   \textbf{a}, Definition of the sphere in momentum space enclosing the Weyl points, and the latitudinal angle $\theta$.  Here, $k_{x,y}$ are given in units of $2\pi/(\sqrt{3}a)$ and $k_z$ is in units of $2\pi/L$ (where $a$ is the unit cell side length and $L$ is the periodicity along $z$, as defined in Fig.~1 of the main text).  For the numerical calculations, the sphere radius is set to $r=0.2$. \textbf{b},\textbf{c}, Calculated evolution of the Berry phase versus $\theta$ for spheres enclosing the Weyl points at $K$ (\textbf{b}) and $H$ (\textbf{c}).  Blue and red indicate the lower and  upper bands respectively.}
\label{fig:charge}
\end{figure}

\section{TIFA modes in a continuum Weyl model}

In the vicinity of a Weyl point, the wavefunctions are governed by an effective long-wavelength Weyl Hamiltonian
\begin{equation}
  \mathcal{H}_0 = -i  \big( \tau_z \sigma_x \partial_x + \sigma_y \partial_y \big)
  +m(k_z) \tau_z\sigma_z,
  \label{H0}
\end{equation}
where $\tau_i$ ($\sigma_i$) denote valley (sublattice) Pauli matrices, and $m(k_z)$ is some function of $k_z$.  Here we rescale the Fermi velocity to 1.  When $k_z \rightarrow 0 (\pi)$, $m(k_z) \rightarrow +(-)k_z$, which indicates the Weyl nodes at $K$ and $K'$ ($H$ and $H'$) have the same chirality. As the system preserves time reversal symmetry, $m(k_z)=-m(-k_z)$. For convenience, here we suppose that $m(k_z)>0$ for $k_z > 0$.

As described in the main text, the TLD is introduced by a ``cut-and-glue'' procedure in which a $\pi/3$ segment is cut from the triangular lattice in the $x$-$y$-plane, and the edges are rejoined.  Ruegg and Lin have previously studied this problem in the context of graphene \cite{Ruegg2013}, and the key steps of their analysis are summarised here for convenience.
  
The effects of the ``cut-and-glue'' procedure are modeled as a boundary condition along the edges of the cut.  From a visual inspection of the lattice, one can deduce that crossing the cut swaps both the valley and sublattice indices.  The boundary condition turns out to have the specific form \cite{Ruegg2013, wang2020}
\begin{equation}
 \Psi_{L} = \begin{pmatrix}
    0 & 0 & 0 & e^{i\eta} \\
    0 & 0 & e^{i\eta^*} & 0 \\
    0 & e^{i\eta^*} & 0 & 0 \\
    e^{i\eta} & 0 & 0 & 0
  \end{pmatrix}\Psi_{U},
  \label{bdcondition}
\end{equation}
where $\eta=2\pi/3$, and $\Psi_{L}$ and $\Psi_{U} $ indicate the wave function for the upper and lower side of the cut, which take the form 
\begin{equation}
    \Psi_{L,U}  = \begin{pmatrix}
    \Psi_{+A} \\ \Psi_{+B} \\ \Psi_{-A} \\ \Psi_{-B}
  \end{pmatrix}_{\!\!L,U}
  \label{wavefunction}
\end{equation}
where $\Psi_{\pm A/B}(r) \equiv \langle r | K_\pm, A/B\rangle$ and $|K_\pm, A/B\rangle$ denote a set of Weyl point states with $\pm$ indexing the Weyl nodes at $K$ and $K'$ (or $H$ and $H'$), and $A/B$ indexing the sublattice. 

To deal with this boundary condition, we introduce polar coordinates $(r, \theta)$ defined in the original (undistorted) space, with $\theta \in [0, 5\pi/3]$. Then we introduce the gauge transformation
\begin{equation}
  H=(SVU)H_0 (SVU)^\dagger, \quad \mathrm{where}\;\;
  \begin{cases}
    U_{\theta} &= e^{i\theta \sigma_{z} \tau_{z}/2} \\
    V_{\theta} &= e^{i\theta \sigma_{y} \tau_{y}/4\Omega} \\
    S &= (1+i\tau_{x} \sigma_{y}), \\
    \Omega &= 5/6.
  \end{cases}
  \label{gauge}
\end{equation}
In polar coordinates, the transformed Hamiltonian is
\begin{equation}
  \mathcal{H}_{\tau'} = \frac{-i}{r}\left[
    \left(r\partial_r+\frac{1}{2}\right) \tau'\sigma_x
    + \left(\partial_{\theta} + i\tau'\frac{1}{4\Omega}\right) \sigma_y
    \right]+ m(k_z) \tau' \sigma_z.
  \label{H1}
\end{equation}
Here, $\tau' = \pm 1$ is a pseudospin index for the two blocks in the transformed Hamiltonian.  The $i(1/4\Omega)\tau'$ term is introduced by the presence of the TLD, and takes the form of a pseudo-magnetic vector potential $\mathbf{A} = (4\Omega r)^{-1} \mathbf{e}_\theta$, corresponding to a singular magnetic flux localised at the origin.

The transformed eigenfunction $\Psi(r,\theta)$ obeys the boundary condition
\begin{equation}
  \Psi(r, \;\theta = 5\pi/3) = -\Psi(r, \theta = 0).
  \label{bcpsi}
\end{equation}
We now define the scaled polar angle $\varphi = \theta/\Omega$, where $\varphi \in [0, 2\pi]$.  The rescaled azimuthal derivative is $\partial_{\varphi} = \partial_{\theta}/\Omega$.  The boundary condition \eqref{bcpsi} now becomes
\begin{equation}
  \Psi(r, \;\varphi = 2\pi) = -\Psi(r, \varphi = 0).
\end{equation}
One can then search for localised solutions by taking a central region of radius $r < \rho$ to be a trivial insulator \cite{Read2000, roy2009, Ruegg2013}. For each $k_z$, there is a single solution arising from one of the two choices of $\tau'$ ($\tau' = 1$ for $0 < k_z < \pi/L$, and $\tau' = -1$ for $-\pi/L < k_z < 0$)\cite{Ruegg2013}; as the system preserve time reversal symmetry, the defect mode for $-k_z$ should be the counterpart of solution at $k_z$ with the same energy but the opposite chirality.

\begin{figure}
\centering
\includegraphics[width=0.4\columnwidth]{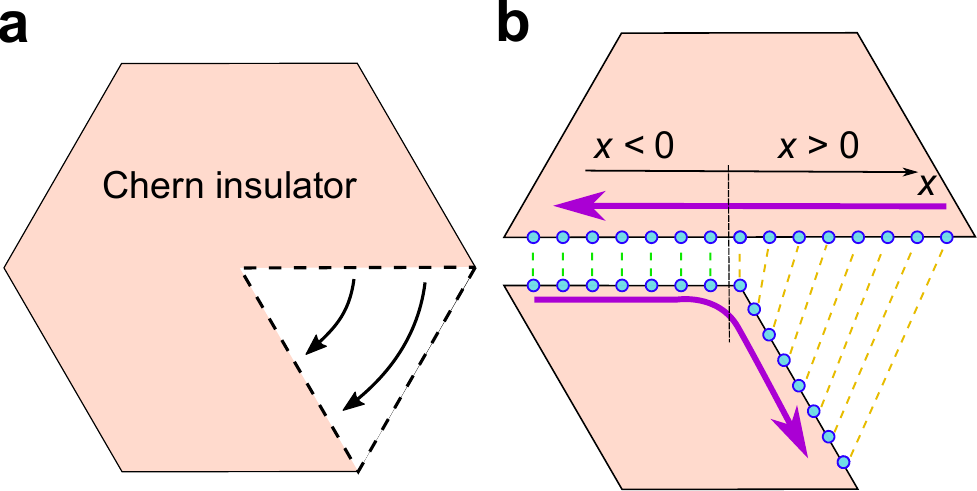}
\caption{Construction of a Chern insulator with a TLD.  \textbf{a}, Cut-and-glue procedure, based on deleting a wedge and reattaching the edges.  \textbf{b}, Procedure based on connecting two differently-shaped semi-infinite domains by adding bonds between the different edges (green and yellow dashes). }
\label{fig:TDs}
\end{figure}

Another way to deduce the existence of the localised TLD-induced modes is to track the evolution of the Chern insulator edge states as the lattice is put together.  Here, we extend an argument from Ref.~\cite{Ruegg2013} to show that the existence of an in-gap defect mode should be insensitive to the details of the lattice configuration near the TLD.

As shown in Fig.~\ref{fig:TDs}\textbf{a}, we are interested in a 2D system formed by a ``cut-and-glue'' construction: a $\pi/6$ wedge is deleted from a Chern insulator, and the edges of the wedges are reattached.  Alternatively, such a system can be assembled as shown in Fig.~\ref{fig:TDs}\textbf{b}: we start with two differently-shaped Chern insulator domains (top and bottom orange regions), and then introduce bonds connecting the different edges (green and yellow dashes in Fig.~\ref{fig:TDs}\textbf{b}).

Prior to the connection, the bulk-edge correspondence principle implies that each domain supports one-way edge states along their boundaries (magenta arrows in Fig.~\ref{fig:TDs}\textbf{b}).  When the connection is established, the edge states couple to each other.  We can treat the parts of the edge far to the left of the TLD ($x \ll 0$ in Fig.~\ref{fig:TDs}\textbf{b}) and far to the right of the TLD ($x \gg 0$ in Fig.~\ref{fig:TDs}\textbf{b}) separately.  Modeling these parts by $L$ and $R$ respectively, we introduce the edge state Hamiltonians
\begin{equation}
  \mathcal{H}_{L,R} = k_x\sigma_z + t\cos{\phi_{L,R}}\; \sigma_x
    + t \sin{\phi_{L,R}}\; \sigma_y.
  \label{H-R}
\end{equation}
Here, $k_x$ is the momentum along the edge, which is a good quantum number far from the TLD; $\sigma_i$ ($i=x,y,z$) are the Pauli matrices, with spinor components representing the two counterpropagating edge states; and $t$ and $\phi_{L,R}$ parameterise an arbitrary coupling between the edge states introduced by the connection.  The phase $\phi_{L,R}$ need not be the same on the two parts of the edge, due to the kink ($60^\circ$ bend) on the lower domain and other local perturbations associated with the central region of the TLD.

In the position representation, Eq.~\eqref{H-R} supports bound solutions of the form
\begin{align}
  \psi_L(x) &\propto \; e^{\sqrt{t^2-E^2}x}
  \begin{pmatrix} te^{-i\phi_L} \\ E+i\sqrt{t^2-E^2} \end{pmatrix}, \\
  \psi_R(x) &\propto \; e^{-\sqrt{t^2-E^2}x}
  \begin{pmatrix} te^{-i\phi_R} \\ E-i\sqrt{t^2-E^2} \end{pmatrix}.
  \label{EV-R}
\end{align}
Now consider continuity relations of the form $\psi_L(0) = e^{i\varphi}\, \psi_R(0)$, where $\varphi$ is an arbitrary phase.  A bound state exists provided
\begin{equation}
  E = t \cos\left(\frac{\phi_L - \phi_R}{2}\right).
\end{equation}
Thus there is a single bound state solution, whose energy ranges from $-t$ to $t$.

\section{Effects of disorder on TIFA modes}

As discussed in the previous section, the TLD-induced Fermi arc (TIFA) modes can be mapped to defect modes of 2D Chern insulators.  This implies that they should be robust against in-plane disorder.  To investigate this, we performed numerical simulations of acoustic structures with varying amounts of disorder.  The disorder is implemented by setting the diameter $d$ of each solid rod passing through the central air sheet as $d = D_0-\Delta \cdot \textrm{rand}(1)$, where $D_0 = 1.6\,\textrm{cm}$ is the rod diameter in the disorder-free case, $\Delta$ parameterises the disorder strength, and $\textrm{rand}(1)$ is drawn independently from a uniform random distribution over $(0,1)$.  The structure depicted in Fig.~\ref{fig:disorder}\textbf{a} was generated with $\Delta=1.2\,\textrm{cm}$ (i.e., rod diameters between $0.4\,\textrm{cm}$ and $1.6\,\textrm{cm}$).  The disorder is assumed to be entirely in-plane, meaning that different layers of the 3D structure have identical disorder configurations.

\begin{figure}
\centering
\includegraphics[width=0.95\columnwidth]{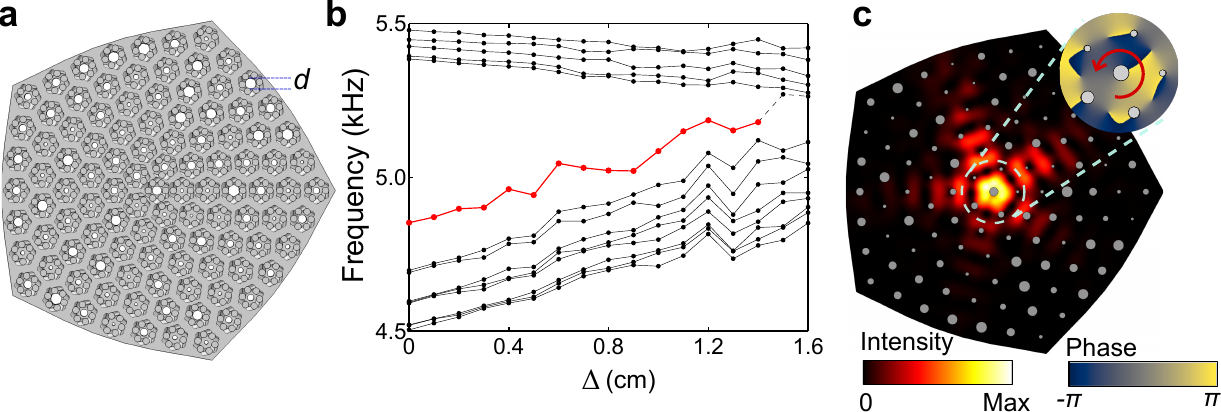}
\caption{Effect of disorder on TIFA modes.  \textbf{a}, Cross-sectional schematic of a disordered structure.  Each solid rod in each unit cell is assigned diameter $\textrm{d} \in [D_0, D_0-\Delta]$ where $D_0 =1.6\,\textrm{cm}$ and $\Delta$ is a disorder strength parameter.  This structure is generated with $\Delta=1.2\,\textrm{cm}$.  All other parameters are the same as in the ordered structure described in the main text.  \textbf{b}, Calculated eigenfrequenies at $k_z = \pi/2L$ for different values of $\Delta$.  Red (black) circles indicate the TIFA modes (bulk modes).  The random rod diameters are re-drawn for each $\Delta$. \textbf{c}, Simulated in-plane acoustic pressure intensity distribution for $\Delta=1.2\,\textrm{cm}$ and $k_z = \pi/2L$, for $z$ located at the midpoint of the central air sheet.  Gray circles are the solid rods with different diameters. Inset: phase distribution near the TLD, showing that the TIFA modes retains its winding number of +1. }
\label{fig:disorder}
\end{figure}

In Fig.~\ref{fig:disorder}\textbf{b}, the eigenfrequencies for the $k_z = \pi/2L$ modes are plotted versus the disorder strength $\Delta$.  With increasing disorder, the bulk gap shrinks, as expected.  The TIFA mode survives up to very strong disorder ($\Delta \approx 1.4\textrm{cm}$) where the gap is almost closed.

Fig.~\ref{fig:disorder}\textbf{c} shows the the acoustic pressure distribution calculated for the TIFA modes at $\Delta=1.2\textrm{cm}$ and  $k_z = \pi/2L$.  (For comparison, refer to the results for the disorder-free structure in Fig.~2\textbf{c} of the main text.)  From the intensity and phase distributions, we see that the mode continues to be localised near the center of the TLD, and that it has OAM of +1, unchanged from the disorder-free case.

\section{Field distribution measurements}

\begin{figure}
\centering
\includegraphics[width=0.95\columnwidth]{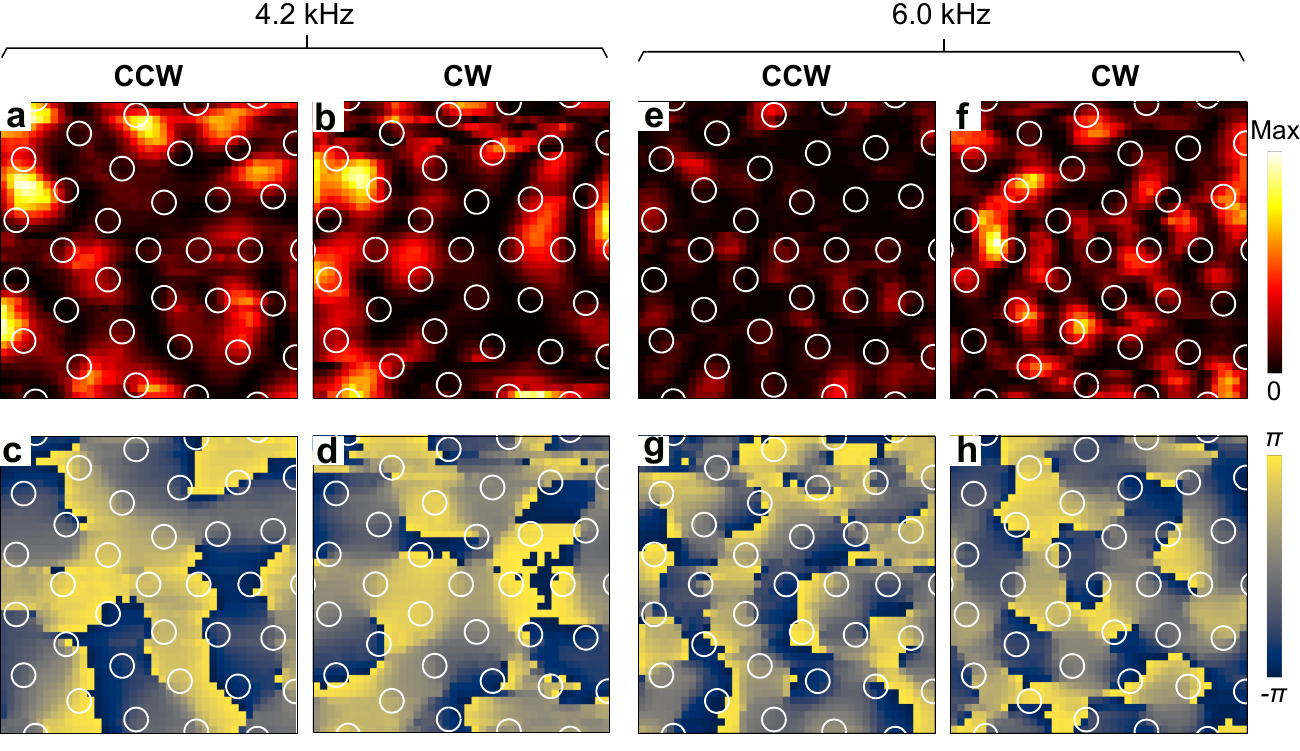}
\caption{Measured field distributions for vortex excitation at frequencies with no TIFA modes.  \textbf{a}--\textbf{d}, Measured intensity distributions (\textbf{a}, \textbf{b}) and phase distributions (\textbf{c}, \textbf{d}) for CCW and CW vortex excitations at $f=4.2\,\textrm{kHz}$.  \textbf{e}--\textbf{h}, Measured intensity distributions (\textbf{e}, \textbf{f}) and phase distributions (\textbf{g}, \textbf{h}) for CCW and CW vortex excitations at $f=6.0\,\textrm{kHz}$.}
\label{fig:outsidesweep2}
\end{figure}

In the main text, we showed the measured field distribution for CW and CCW vortex excitation at $5.6\,\textrm{kHz}$, a frequency that supports TIFA modes (Fig.~4\textbf{c}--\text{h}).  For comparison, Fig.~\ref{fig:outsidesweep2} shows the measured field distributions at $4.2\,\textrm{kHz}$ and $6.0\,\textrm{kHz}$, frequencies outside the range where TIFA modes are predicted to exist.  In these two cases, neither vortex produces a localised hot spot near the center of the TLD.

\begin{figure}
\centering
\includegraphics[width=0.4\columnwidth]{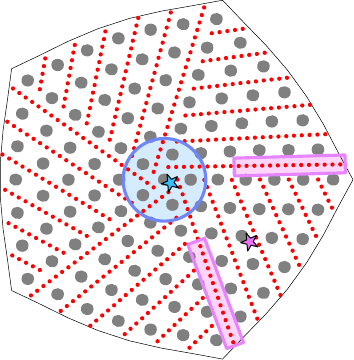}
\caption{Probe and source positions in each sample layer.  Red circles indicate the probe positions for field-mapping measurements.  Blue and magenta stars indicate the position of the source for defect and bulk excitation, respectively.  Blue circle and magenta rectangles indicate the measurement points used to plot the defect and bulk spectra, respectively. }
\label{fig:Measuredpoints}
\end{figure}

Fig.~\ref{fig:Measuredpoints} shows the source and probe positions inside the sample layers for the various experiments.  The probe positions used for field mapping (Fig.~3\textbf{d}--\textbf{f} and Fig.~4\textbf{c},\textbf{d} of the main text) are shown as red circles.  When exciting close to the defect (Fig.~3\textbf{a} and Fig.~3\textbf{c}--\textbf{f} of the main text), the source is located at the blue star in the bottom layer of the sample.  When exciting the bulk, the source is located at the magenta star in the bottom layer.  When averaging the spectra over points close to the TLD (Fig.~3\textbf{a} and Fig.~4\textbf{b} of the main text), we use the points enclosed in the blue circle.  For bulk spectra (Fig.~3\textbf{b} of the main text), we average the intensities over the points in the magenta rectangles, away from the TLD.

\clearpage


\end{widetext}
\end{document}